\documentclass[times,watermark]{zHenriquesLab-StyleBioRxiv}
\usepackage{blindtext}

\usepackage{graphicx}
\usepackage{amsmath}
\usepackage[version=4]{mhchem}
\usepackage{siunitx}
\usepackage{longtable,tabularx}
\usepackage{xcolor}
\usepackage{xfrac}
\usepackage{amsmath}
\usepackage{color,soul}
\DeclareUnicodeCharacter{2212}{-}
\usepackage{dcolumn}
\usepackage{bm}
\usepackage[utf8]{inputenc}
\usepackage[T1]{fontenc}
\usepackage{mathptmx}

\leadauthor{Tancin} 

\begin{document}

\title{Ultrafast-laser-absorption spectroscopy for single-shot, mid-infrared measurements of temperature, CO, and CH$_4$ in flames}

\shorttitle{Ultrafast-laser-absorption spectroscopy in the mid-IR}

\author[1,*]{Ryan J. Tancin}
\author[2]{Ziqiao Chang}
\author[2]{Mingming Gu}
\author[2]{Vishnu Radhakrishna}
\author[2]{Robert P. Lucht}
\author[2]{Christopher S. Goldenstein}

\affil[1]{School of Aeronautics and Astronautics, Purdue University, 585 Purdue Mall, West Lafayette, IN 47907, USA}
\affil[2]{School of Mechanical Engineering, Purdue University, 585 Purdue Mall, West Lafayette, Indiana 47907, USA}
\affil[*]{Corresponding author: rtancin@purdue.edu}

\maketitle

\begin{abstract}
This manuscript describes the development of an ultrafast (i.e., femtosecond), mid-infrared, laser-absorption diagnostic and its initial application to measuring temperature, CO, and CH$_4$ in flames. The diagnostic employs a Ti:Sapphire oscillator emitting 55-fs pulses near 800 nm which were amplified and converted into the mid-infrared (mid-IR) though optical parametric amplification (OPA) at a repetition rate of 5 kHz. The pulses were directed through the test gas and into a high-speed mid-infrared spectrograph to image spectra across a $\approx$30 nm bandwidth with a spectral resolution of $\approx$0.3 nm. Gas properties were determined by least-squares fitting a spectroscopic model to measured single-shot absorbance spectra. The diagnostic was validated with measurements of temperature, CO, and CH$_4$ in a static-gas cell with an accuracy of 0.7\% to 1.8\% of known values. Single-shot, 5 kHz measurements of temperature and CO were acquired near 4.9 $\mu$m in a laser-ignited HMX (i.e., 1,3,5,7-tetranitro-1,3,5,7-tetrazoctane) flame and exhibited a 1-$\sigma$ precision of 0.4\% at $\approx$2700 K. Further, CH$_4$ and temperature measurements were acquired near 3.3 $\mu$m in a partially premixed CH$_4$-air flame produced by a Hencken burner and exhibited a precision of 0.3\% at $\approx$1000 K.
\end{abstract}



\setlength{\parindent}{2em}

\section*{Introduction}

Laser-absorption spectroscopy (LAS) is a powerful technique for non-invasive, quantitative measurements of temperature and species concentrations in combustion environments \cite{Goldenstein2017b}. LAS diagnostics often employ narrowband, wavelength-tunable lasers (e.g., tunable diode lasers, quantum-cascade lasers) which are capable of measuring gas conditions via spectra measured over several cm$^{-1}$ at rates up to 1 MHz \cite{Goldenstein2017b, Pineda2019}. While highly useful, the narrowband nature of this approach can: 1) limit the dynamic range of such diagnostics \cite{An2011a}, 2) complicate measurements of molecules with broad spectra (e.g., at high pressures), and 3) often prevents multi-species measurements using a single light source. To address these issues, numerous researchers have developed broadband LAS diagnostics, typically with 10s to 1000s of cm$^{-1}$ of spectral bandwidth \cite{Sanders2001, Rein2017, Strand2019a, Sanders2002, GoranBlume2015, Schroeder2017a, Draper2019, Stauffer2019a}.

\begin{figure*}[htbp]
\centering
\includegraphics[width=\textwidth]{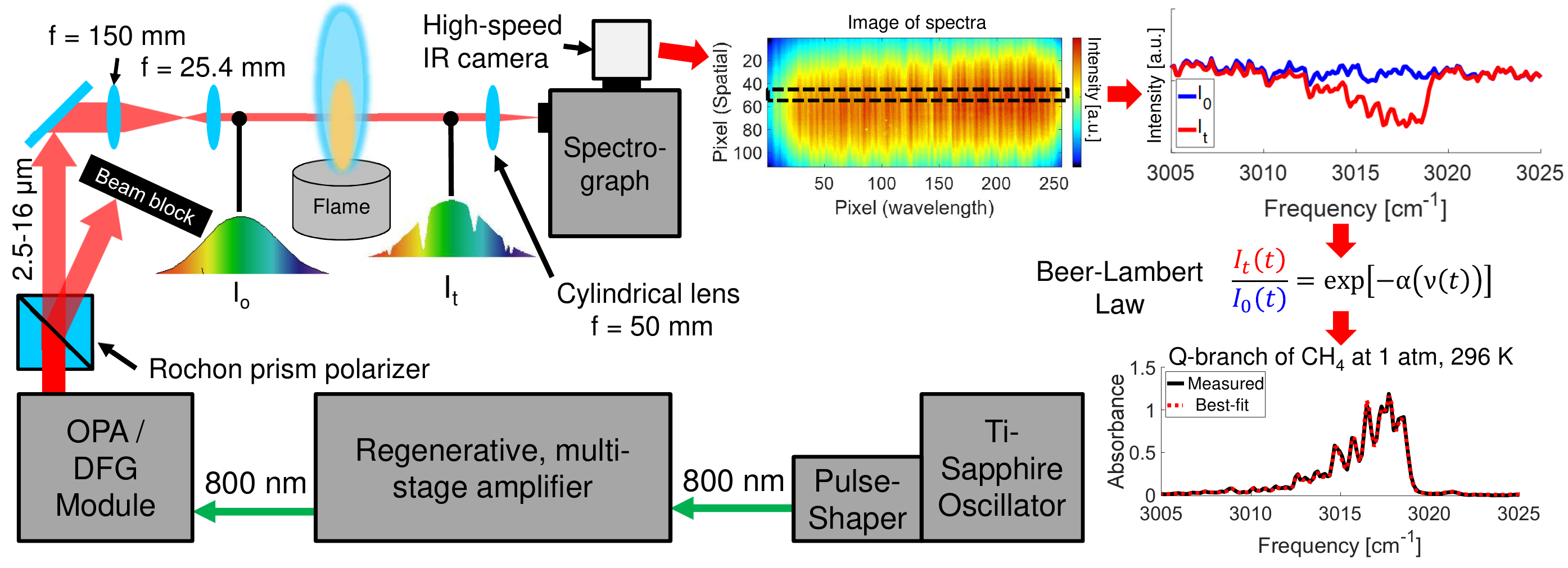}
\caption{Concept schematic illustrating the experimental setup and data processing procedure employed by the ultrafast-laser-absorption diagnostic. }
\label{fig:Setup}
\end{figure*}

One common broadband LAS technique employs scanned-wavelength lasers with broad wavelength-tuning capabilities. For example, Sanders et al. \cite{Sanders2001} used a vertical-cavity surface-emitting laser (VCSEL) to scan $\approx$30 cm$^{-1}$ near 760 nm in order to acquire measurements of temperature and O$_2$ at 500 Hz in high-pressure gases. More recently, a MEMS-VCSEL capable of tuning $\approx$170 cm$^{-1}$ near 1.4 $\mu$m has been used to measure temperature and H$_2$O at 100 kHz in a wide variety of combustion applications (see \cite{Rein2017} and references therein). In addition, external-cavity quantum-cascade lasers in the mid- and far-IR with large-amplitude ($\approx$100 cm$^{-1}$) tuning have emerged, which recently enabled measurements of C$_2$H$_4$ between 8.5 and 11.7 $\mu$m in shock-heated gases with a time resolution of 3 ms \cite{Strand2019a}. Unfortunately, light sources suitable for this approach are limited to only a few wavelength regions and the time-resolution is limited to the reciprocal of the scan frequency (typically on the order of 1 ms to 10 $\mu$s).

Numerous techniques employing ultrafast lasers (i.e., lasers emitting ultrashort, typically < 10 ps, pulses) have also been used for broadband LAS measurements of combustion gases, however they have not been applied in the mid-infrared. For example, Sanders \cite{Sanders2002} used a femtosecond-fiber-laser-pumped supercontinuum source and a scanning bandpass filter to produce 1000 cm$^{-1}$ of bandwidth near 1450 nm for measurements of temperature and several species (H$_2$O, CO$_2$, C$_2$H$_2$, C$_2$H$_6$O) at 50 kHz. More recently, Blume and Wagner \cite{GoranBlume2015} used a commercial supercontinuum lightsource with a dispersing fiber to acquire measurements of CH$_4$ and temperature at 200 Hz using 110 cm$^{-1}$ of bandwidth near 1650 nm. One drawback to this approach is the relatively large noise levels inherent to supercontinuum lightsources which has prevented single-shot measurements. Recently, Draper et al. \cite{Draper2019} applied a dual-frequency-comb spectrometer to characterize combustion environments. The authors used 160 cm$^{-1}$ of bandwidth near 1655 nm to measure temperature and CH$_4$ concentration in a rapid compression machine. Currently, multi-shot averaging and filtering have extended the repetition rate to $\approx$1.4 kHz \cite{Draper2019}. Most recently, time-resolved, optically gated absorption (TOGA) spectroscopy was developed by Stauffer et. al. \cite{Stauffer2019a} to provide background-free measurements of absorption spectra. The authors used an amplified Ti:Sapphire laser and two frequency-doubling processes to produce 100-fs pulses centered near 310 nm with over 300 cm$^{-1}$ of bandwidth to acquire spectra of OH.

\begin{figure}[!b]
\centering
\includegraphics[width=\linewidth]{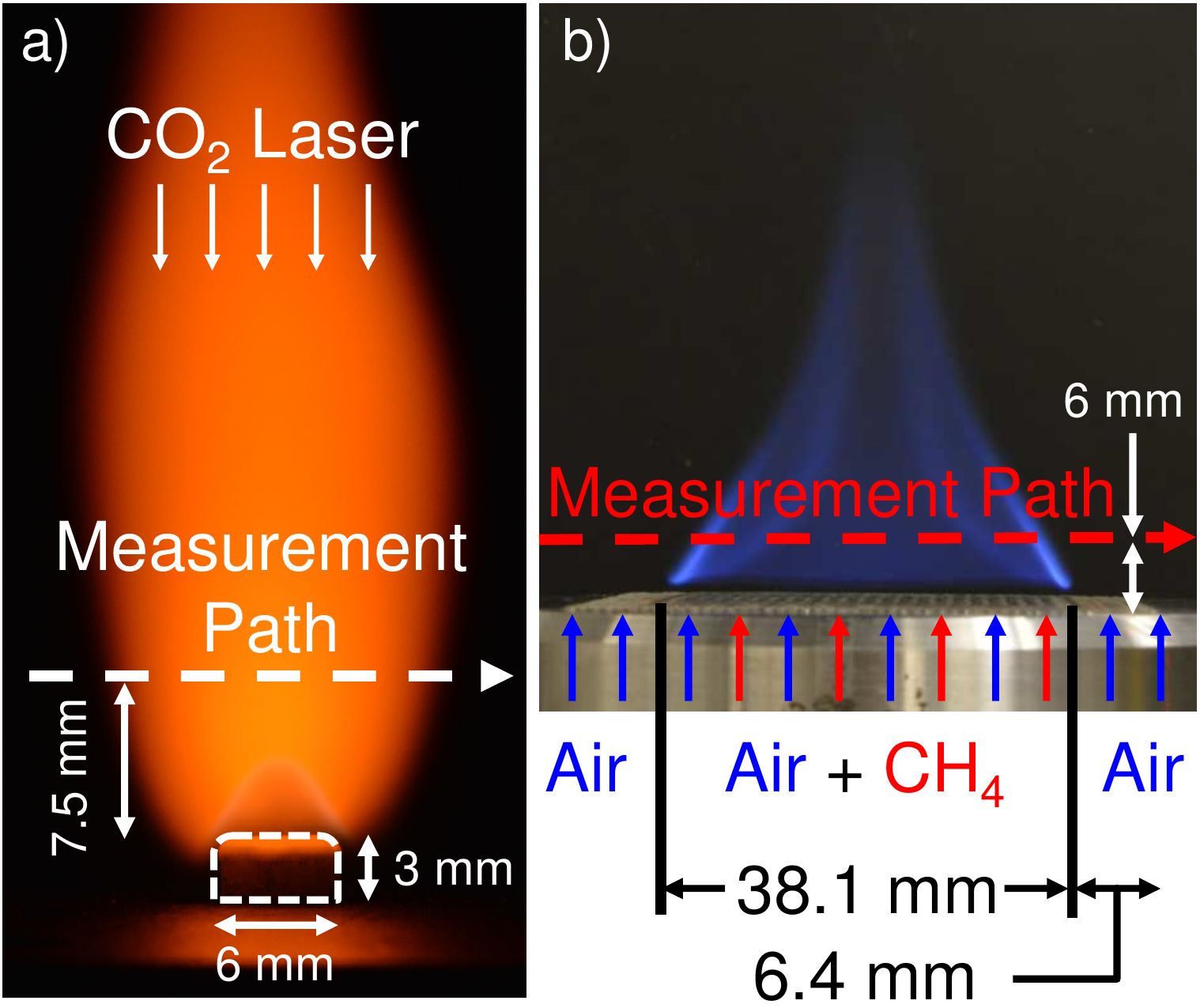}
\caption{Images of and pertinent dimensions for the a) laser-ignited HMX flame and b) partially premixed CH$_4$-air flame studied here.}
\label{fig:Flame Image}
\end{figure}

Here we demonstrate the first, to our knowledge, ultrafast, single-shot, mid-infrared LAS diagnostic for temperature and species measurements in combustion gases. In addition to the aforementioned potential advantages of broadband LAS diagnostics (e.g., improved high-pressure capability, high-dynamic range, multi-species measurements), we demonstrate that this diagnostic offers several unique advantages: 1) ultrafast (<1 ps) time resolution, 2) access to strong fundamental absorption bands located throughout the mid-wave infrared (2 to 5.5 $\mu$m) using a single light source and camera, and 3) potential for single-shot, spatially resolved (1D) thermometry and species measurements at 5 kHz. This manuscript describes the design and operating principles of this diagnostic technique, in addition to its initial validation and application to characterizing flames. Specifically, single-shot, 5 kHz measurements of temperature and CO were acquired in a laser-ignited HMX flame using wavelengths near 4.9 $\mu$m, and measurements of temperature and CH$_4$ were acquired in a partially premixed CH$_4$-air Hencken-burner flame using wavelengths near 3.3 $\mu$m. The results demonstrate that this diagnostic is capable of providing high-precision measurements of temperature (<1\%) and molecular species (<2.5\%) in combustion gases with ultrafast (<1 ps) time resolution.

\begin{figure*}[!t]
\centering
\includegraphics[width=1\textwidth]{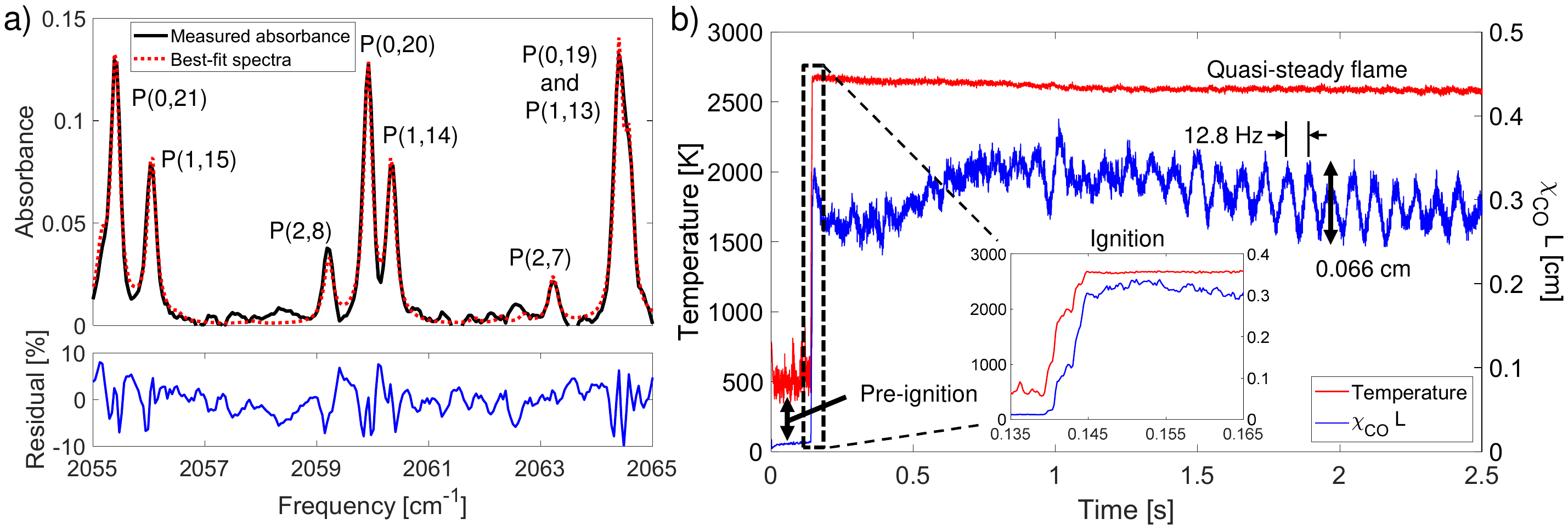}
\caption{Example single-shot measurement and best-fit spectrum for CO near 4.9 $\mu$m  (a) and corresponding time history of temperature and $\chi_{CO}L$ acquired in a laser-ignited HMX flame (b).}
\label{fig:CO_data}
\end{figure*}

\section*{Experimental Procedure}

The experimental system and data processing procedures are summarized in Fig. \ref{fig:Setup}. Ultrafast laser pulses were generated by a mode-locked Ti:Sapphire laser (Coherent Mantis) at a rate of 80 MHz and a center wavelength near 800 nm. The pulses were tailored by a Femtojock pulse-shaper to prepare them for amplification by a multi-stage regenerative chirped-pulse amplifier (Coherent Legend Elite Duo). The amplifier produced pulses with a pulse energy of 2 mJ and duration of 55 fs at a repetition rate of 5 kHz. Next, the pulses were directed into an optical parametric amplifier (OPA) module (Coherent OPerA Solo) equipped with non-collinear difference-frequency generation (NDFG) crystals to generate ultrashort mid-IR pulses. The mid-IR pulses contain $\approx$600 cm$^{-1}$ of useful spectral bandwidth and the center wavelength can be tuned between 2.5 and 18 $\mu$m via computer controlled manipulation of the crystals. For the results presented here, the pulse energies were 30 $\mu$J and 4 $\mu$J for wavelengths near 3315 nm and 4858 nm, respectively. A MgF$_2$ Rochon prism polarizer was used to attenuate the pulse energy to $\approx$0.75 $\mu$J to prevent camera saturation. Next, the mid-IR pulses were directed through the test gas and then focused onto the slit of an Andor Shamrock 500i imaging spectrometer using a cylindrical lens (ZnSe, 50 mm focal length). A 300 groove/mm diffraction grating was used to spectrally disperse the pulses and a Telops FAST-IR 2k high-speed IR camera was used to image each individual pulse in 2D. This configuration provided a theoretical spectral resolution and bandwidth of 0.21 nm and 38 nm, respectively, and similar metrics were achieved in practice. The IR-camera recorded 52x256 pixels with a 5 $\mu$s exposure time at 10k frames-per-second in order to record two images per pulse (one for the pulse, another to image background emission between pulses). The camera was synchronized with the laser output. 

CO and CH$_4$ were chosen for the initial application of this diagnostic primarily to demonstrate its ability to measure gas properties via molecules with well isolated lines (CO) or blended spectra (CH$_4$). Measurements were acquired in CO's P-branch near 2060 cm$^{-1}$ due to favorable absorbance levels, lack of interference from other combustion relevant species (e.g., H$_2$O, CO$_2$), and previous combustion studies demonstrating the utility of this spectral region \cite{Spearrin2014a,Lee2018,Tancin2019b, Tancin2019c}. Temperature and CH$_4$ measurements were acquired via CH$_4$'s Q-branch near 3015 cm$^{-1}$ due to its strength, isolation from interfering absorption lines, and the large number of transitions with unique lower-state energy comprising it which enhances temperature sensitivity \cite{Goldenstein2017b,An2011a}.

\begin{figure*}[!t]
\centering
\includegraphics[width=1\textwidth]{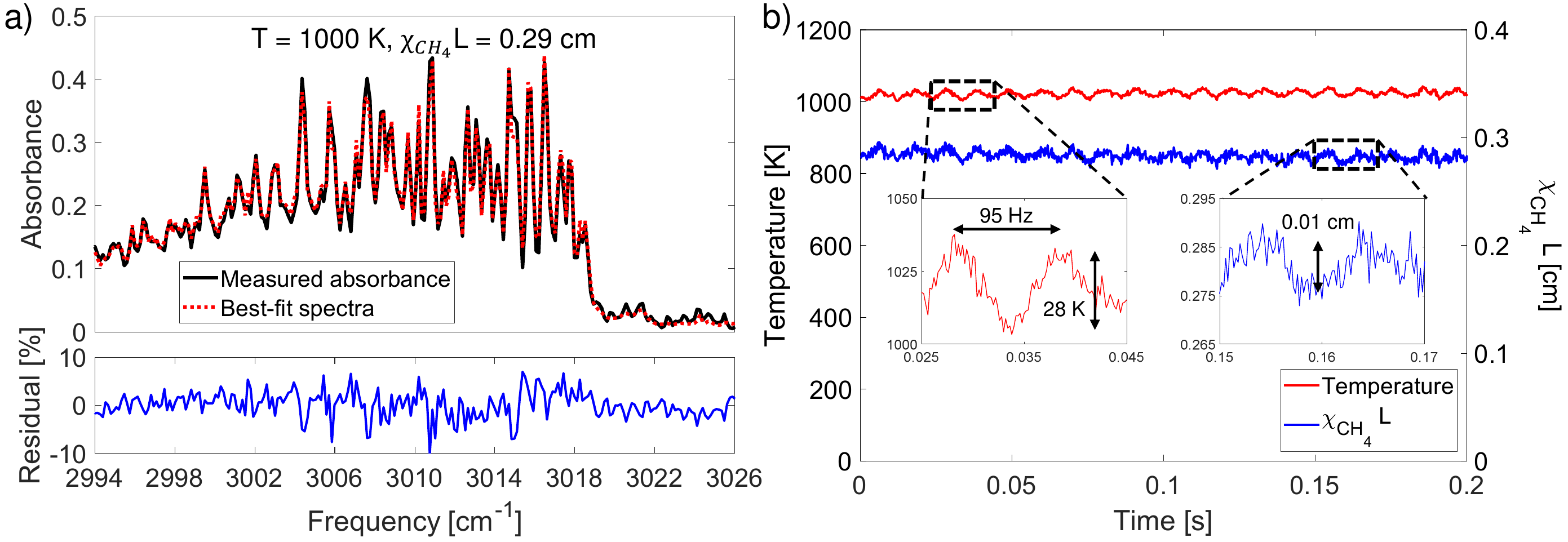}
\caption{Example single-shot measurement and best-fit spectrum of CH$_4$'s Q-branch (a) and corresponding time history of temperature and $\chi_{CH_4}L$ acquired in partially premixed CH$_4$-air flame (b).}
\label{fig:CH4_data}
\end{figure*}

The center wavelength of the spectrograph and light source were set to 4858 nm for measurements of temperature and CO concentration. Measurements were acquired in a static-gas cell \cite{Schwarm2019} (9.4 cm path length) filled with a commercially prepared gas mixture (2\% CO, 1.8\% CO$_2$ and 96.2\% N$_2$ by mole) at 1 atm and 296 K to validate the accuracy of the diagnostic and data processing routine. Measurements of temperature and CO column density ($\chi_{CO}L$) were also acquired in a laser-ignited HMX flame (see Figure \ref{fig:Flame Image}a). The HMX was pressed into a 6 mm diameter, 3 mm tall cylindrical pellet and ignited with a CO$_2$ laser emitting 78 W/cm$^2$ at a wavelength of 10.6 $\mu$m for the entire test duration. For this test, the 1/e$^2$ diameter of the ultrafast beam was reduced from 7.9 mm to 1.4 mm via a lens telescope consisting of a CaF$_2$ plano-convex lens ($f$=150 mm) and an AR-coated Si plano-convex lens ($f$=25.4 mm). The measurement was acquired 7.5 mm above the initial surface location of the pellet.

The accuracy of the temperature and CH$_4$ diagnostic was also validated via measurements in a static-gas cell (0.7 cm path length) and Figure \ref{fig:Setup} illustrates an example single-shot measurement of CH$_4$'s Q-branch at 296 K and 1 atm. The cell was filled with a commercially prepared  mixture of 7\% CH$_4$, 15\% C$_2$H$_2$ and 78\% N$_2$ by mole at 1 atm and 296 K. Measurements were also acquired in a laminar, partially-premixed CH$_4$-air flame (see Figure \ref{fig:Flame Image}b). The flame was produced by a Hencken burner with a 38.1 mm square core consisting of a honeycomb of alternating jets of CH$_4$ and air surrounded by a 6.4 mm thick co-flow of air. The core-flow equivalence ratio was 6.4 with an exit plane velocity of $\approx$0.9 cm/s. 

For each experiment, the measured spectral absorbance ($\alpha$) was calculated at each frequency ($\nu$) using the measured incident ($I_o$) and transmitted ($I_t$) laser intensity and Beer's Law: $\alpha(\nu) = -ln[I_t(\nu)/I_o(\nu)]$. $I_o$ was obtained empirically by recording and averaging (to reduce noise) 200 images of ultrashort mid-IR pulses in the absence of the test gas. The signal-to-noise ratio (SNR) of $I_o$ was then further increased by averaging 8 to 18 (depending on test) spectra of $I_o$ (acquired by adjacent rows of pixels within a single image) together. For the results reported here, this corresponds to spatial averaging across less than 1 mm (vertically) at the measurement location in the flame gas.

The gas temperature and absorbing-species concentration were determined by least-squares fitting a simulated absorbance spectra to the measured absorbance spectra. The fitting routine employs the following free parameters: temperature, absorbing species concentration, a frequency shift, the width of the instrument response function (IRF), and a scaling factor for $I_o$ to account for shot-to-shot fluctuations in pulse energy. The frequency axis of the data was determined by matching 3 to 4 prominent absorption features to their linecenter frequencies (provided by a spectroscopic database) and fitting a linear frequency axis to those points. Absorbance spectra were simulated in the fitting routine using the following procedure: 1) A high-resolution ($\approx$0.0005 cm$^{-1}$) absorbance spectrum was simulated using $\alpha_{HR}(\nu)= \sum_jS_j(T)\phi_j(\nu)P\chi L$ and a spectroscopic database where $S_j$ and $\phi_j$ are the linestrength and lineshape of transition $j$, respectively, $T$ is the gas temperature, $P$ is the gas pressure, $\chi$ is the absorbing species mole fraction and $L$ is the path length through the absorbing gas. Transition lineshapes were modeled as a Voigt profile. 2) The resolution of the empirical baseline was increased to match that of $\alpha_{HR}$ via linear interpolation, thereby yielding $I_{o,HR}$. 3) Next, Eq. \ref{IRF convolution:equation} was used to produce a high-resolution, semi-empirical transmission spectrum ($I_{t,HR,conv}$), which has been convolved with the IRF of the spectrograph-camera setup. 4) $I_{o,HR}$ was then convolved with the IRF to yield $I_{o,HR,conv}$. 5) A simulated absorbance spectrum with the IRF accounted for was then calculated from $I_{o,HR,conv}$ and $I_{t,HR,conv}$ using Beer's Law and then downsampled to the same frequency axis as the measured absorbance spectrum to enable direct comparison and the sum-of-squared error (the optimization parameter) to be computed by the fitting routine.
\begin{multline}
\label{IRF convolution:equation}
I_{t,HR,conv}(\lambda) = \int_{-\infty}^{\infty} IRF(\lambda-\tau)\times\\ 
e^{-\alpha_{HR}(T,P,\chi,L)} d\tau
\end{multline}

\noindent Here, $\lambda$ is the wavelength and $\tau$ is the convolution shift variable. The IRF was modeled as a Lorentzian lineshape.

It is important to note that the IRF is not directly convolved with the simulated absorbance spectrum because the spectrograph-camera setup "sees" a transmission spectrum, not an absorbance spectrum. The convolution with the IRF was executed in wavelength-space where the full-width at half maximum (FWHM) of the IRF is constant. The spectroscopic model for CO employed the HITEMP 2019 database \cite{Hargreaves2019} and the spectroscopic model for CH$_4$ utilized a preliminary version (yet to be openly released) of the HITEMP 2019 database for methane \cite{Hargreaves2019} which is based on work by Rey et. al. \cite{Rey2017}.

\section*{Results and Discussion}

Measurements of temperature and CO concentration in static-gas cell experiments were accurate within 1.5\% and 1.8\% of known values, respectively, with a 1-$\sigma$ precision of 0.7\% and 1.2\%, respectively. Figure \ref{fig:CO_data}a displays a representative measured and best-fit absorbance spectra acquired in the HMX flame during quasi-steady-state. The 1-$\sigma$ noise level in the measured absorbance was 0.25\%. Time histories of temperature and $\chi_{CO} L$ are shown in Figure \ref{fig:CO_data}b. The diagnostic resolved pre-ignition decomposition of HMX through quasi-steady burning. At quasi-steady-state, the measured time-averaged temperature and CO column density was 2620 K and 0.32 cm (i.e., 21.3\% by mole for a 1.5 cm thick flame) respectively, which agree well with measurements acquired in our lab using quantum-cascade lasers similar to as reported in \cite{Tancin2019b}. The 1-$\sigma$ precisions are 10 K (0.4\%) and 0.007 cm (2.3\%). Throughout the burn, the temperature decreased by $\approx$100 K due to the surface of HMX regressing away from the measurement line-of-sight. Large amplitude oscillations in $\chi_{CO} L$ are observed at 12.8 Hz as a result of a natural flame instability \cite{Finlinson1994}. Temperature measurements acquired prior to ignition were enabled by the broad spectral bandwidth of the measurement. Measurements with an absorbance SNR as low as 20 were made, which occurred at a CO column density of 0.003 cm and temperature of 500 K. 


The temperature and CH$_4$ concentration measured in static-gas cell experiments were accurate within 1.2\% and 0.8\% of known values, respectively, with a 1-$\sigma$ precision of 0.2\% and 0.7\%, respectively. Figure \ref{fig:CH4_data}a shows a typical single-shot absorbance spectra of CH$_4$'s Q-branch and Figure \ref{fig:CH4_data}b shows measured time histories of temperature and CH$_4$ column density acquired in a partially-premixed CH$_4$-air Hencken-burner flame. The time-averaged values of temperature and column density are 1023 K and 0.283 cm, both of which exhibit a 95 Hz oscillation (due to a combustion instability) with a peak-to-peak amplitude of 28 K and 0.01 cm, respectively. After accounting for this oscillation, the 1-$\sigma$ precision of the measured temperature and column density are 3 K (0.3\%) and 0.003 cm (1\%), respectively. The results shown were acquired with an absorbance-noise level of $\approx$ 0.5\%. 

The results presented in this manuscript illustrate that ultrafast-laser-absorption spectroscopy in the mid-infrared is capable of high-fidelity characterization of combustion gases with ultrafast time resolution. 

\section*{Acknowledgements}
This work was supported by AFOSR Grant FA9550-18-1-0210 with Dr. Mitat Birkan as program manager and NSF CBET Grant 1834972. The authors thank Austin McDonald for advice regarding the development of the spectral-fitting routine and Dr. Robert Hargreaves for sharing a preliminary version of the HITEMP 2019 database for CH$_4$.

\section*{References}
\bibliography{ULAS_Paper_1}

\begin{thebibliography}{10}
\newcommand{\enquote}[1]{``#1''}

\bibitem{Goldenstein2017b}
C.~S. Goldenstein, R.~M. Spearrin, J.~B. Jeffries, and R.~K. Hanson,
  \enquote{{Infrared laser-absorption sensing for combustion gases},} Progress
  in Energy and Combustion Science \textbf{60}, 132--176 (2017).

\bibitem{Pineda2019}
D.~I. Pineda, F.~A. Bendana, K.~K. Schwarm, and R.~M. Spearrin,
  \enquote{{Multi-isotopologue laser absorption spectroscopy of carbon monoxide
  for high-temperature chemical kinetic studies of fuel mixtures},} Combustion
  and Flame \textbf{207}, 379--390 (2019).

\bibitem{An2011a}
X.~An, A.~W. Caswell, and S.~T. Sanders, \enquote{{Quantifying the temperature
  sensitivity of practical spectra using a new spectroscopic quantity:
  Frequency-dependent lower-state energy},} Journal of Quantitative
  Spectroscopy and Radiative Transfer \textbf{112}, 779--785 (2011).

\bibitem{Sanders2001}
S.~T. Sanders, J.~Wang, J.~B. Jeffries, and R.~K. Hanson, \enquote{Diode-laser
  absorption sensor for line-of-sight gas temperature distributions,} Applied
  Optics. \textbf{40}, 4404--4415 (2001).

\bibitem{Rein2017}
K.~D. Rein, S.~Roy, S.~T. Sanders, A.~W. Caswell, F.~R. Schauer, and J.~R.
  Gord, \enquote{Measurements of gas temperatures at 100 khz within the annulus
  of a rotating detonation engine,} Applied Physics B: Lasers and Optics
  \textbf{123}, 88 (2017).

\bibitem{Strand2019a}
C.~L. Strand, Y.~Ding, S.~E. Johnson, and R.~K. Hanson, \enquote{{Measurement
  of the mid-infrared absorption spectra of ethylene (C$_2$H$_4$) and other
  molecules at high temperatures and pressures},} Journal of Quantitative
  Spectroscopy and Radiative Transfer \textbf{222-223}, 122--129 (2019).

\bibitem{Sanders2002}
S.~T. Sanders, \enquote{{Wavelength-agile fiber laser using group-velocity
  dispersion of pulsed super-continua and application to broadband absorption
  spectroscopy},} Applied Physics B: Lasers and Optics \textbf{75}, 799--802
  (2002).

\bibitem{GoranBlume2015}
N.~{G{\"{o}}ran Blume} and S.~Wagner, \enquote{{Broadband supercontinuum laser
  absorption spectrometer for multiparameter gas phase combustion
  diagnostics},} Optics Letters \textbf{40}, 3141 (2015).

\bibitem{Schroeder2017a}
P.~J. Schroeder, R.~J. Wright, S.~Coburn, B.~Sodergren, K.~C. Cossel,
  S.~Droste, G.~W. Truong, E.~Baumann, F.~R. Giorgetta, I.~Coddington, N.~R.
  Newbury, and G.~B. Rieker, \enquote{{Dual frequency comb laser absorption
  spectroscopy in a 16 MW gas turbine exhaust},} Proceedings of the Combustion
  Institute \textbf{36}, 4565--4573 (2017).

\bibitem{Draper2019}
A.~D. Draper, R.~K. Cole, A.~S. Makowiecki, A.~Zdanawicz, J.~Mohr, A.~Marchese,
  N.~Hoghoogi, and G.~B. Rieker, \enquote{{Progress toward dual frequency comb
  spectroscopy in a rapid compression machine},} Optics Express \textbf{27},
  10814--10825 (2019).

\bibitem{Stauffer2019a}
H.~U. Stauffer, P.~S. Walsh, S.~Roy, and J.~R. Gord, \enquote{{Time-resolved
  optically gated absorption (TOGA) spectroscopy: A background-free,
  single-shot broadband absorption method for combusting flows},} AIAA Scitech
  2019 Forum p. 1607 (2019).

\bibitem{Spearrin2014a}
R.~M. Spearrin, C.~S. Goldenstein, I.~A. Schultz, J.~B. Jeffries, and R.~K.
  Hanson, \enquote{{Simultaneous sensing of temperature, CO, and CO$_2$ in a
  scramjet combustor using quantum cascade laser absorption spectroscopy},}
  Applied Physics B: Lasers and Optics \textbf{117}, 689--698 (2014).

\bibitem{Lee2018}
D.~D. Lee, F.~A. Bendana, S.~A. Schumaker, and R.~M. Spearrin,
  \enquote{Wavelength modulation spectroscopy near 5 $\mu$m for carbon monoxide
  sensing in a high-pressure kerosene-fueled liquid rocket combustor,} Applied
  Physics B: Lasers and Optics \textbf{124}, 77 (2018).

\bibitem{Tancin2019b}
R.~J. Tancin, G.~C. Mathews, and C.~S. Goldenstein, \enquote{Design and
  application of a high-pressure combustion chamber for studying propellant
  flames with laser diagnostics,} Review of Scientific Instruments \textbf{90},
  045111 (2019).

\bibitem{Tancin2019c}
R.~J. Tancin, R.~M. Spearrin, and C.~S. Goldenstein, \enquote{{2D mid-infrared
  laser-absorption imaging for tomographic reconstruction of temperature and
  carbon monoxide in laminar flames},} Optics Express \textbf{27}, 14184
  (2019).

\bibitem{Schwarm2019}
K.~K. Schwarm, H.~Q. Dinh, C.~S. Goldenstein, D.~I. Pineda, and R.~M. Spearrin,
  \enquote{{High-pressure and high-temperature gas cell for absorption
  spectroscopy studies at wavelengths up to 8 µm},} Journal of Quantitative
  Spectroscopy and Radiative Transfer \textbf{227}, 145--151 (2019).

\bibitem{Hargreaves2019}
R.~J. Hargreaves, I.~E. Gordon, L.~S. Rothman, S.~A. Tashkun, V.~I. Perevalov,
  A.~A. Lukashevskaya, S.~N. Yurchenko, J.~Tennyson, and H.~S. M{\"{u}}ller,
  \enquote{{Spectroscopic line parameters of NO, NO$_2$, and N$_2$O for the
  HITEMP database},} Journal of Quantitative Spectroscopy and Radiative
  Transfer \textbf{232}, 35--53 (2019).

\bibitem{Rey2017}
M.~Rey, A.~V. Nikitin, and V.~G. Tyuterev, \enquote{{Accurate theoretical
  methane line lists in the infrared up to 3000 K and quasi-continuum
  absorption/emission modeling for astrophysical applications},} The
  Astrophysical Journal \textbf{847}, 105 (2017).

\bibitem{Finlinson1994}
J.~C. Finlinson, T.~Parr, and D.~Hanson-Parr, \enquote{{Laser recoil, plume
  emission, and flame height combustion response of HMX and RDX at atmospheric
  pressure},} Symposium (International) on Combustion \textbf{25}, 1645--1650
  (1994).

\end{thebibliography}


\end{document}